\documentclass[aps,prl,twocolumn,showpacs,groupedaddress]{revtex4}

\usepackage{graphicx}
\graphicspath{{pict/}{}}
\usepackage{bm}

\newcommand\pictc[5]{\begin{figure}
            \centerline{\vspace{0mm}
\includegraphics[width=#1\columnwidth,height=0.7\textheight,keepaspectratio]{#3}}
            \protect\caption{\protect\label{fig:#4} #5}
                    \end{figure}            }

\newcommand\pict[4][1.0]{\pictc{#1}{!tb}{#2}{#3}{#4}}
\newcommand\rpict[1]{\ref{fig:#1}}

\newcommand\leqt[1]{\protect\label{eq:#1}}
\newcommand\reqtn[1]{\ref{eq:#1}}
\newcommand\reqt[1]{(\reqtn{#1})}

\newcounter{Fig}

\begin{document}
\begin{sloppy}

\title{Broadband diffraction management and
self-collimation of white light in photonic lattices}

\author{Ivan L. Garanovich}
\author{Andrey A. Sukhorukov}
\author{Yuri S. Kivshar}

\affiliation{Nonlinear Physics Centre and Centre for Ultra-high bandwidth
Devices for Optical Systems (CUDOS),
 Research School of Physical Sciences and Engineering,
 Australian National University, Canberra, ACT 0200, Australia}

\begin{abstract}
We suggest a novel type of photonic structures where the strength of
diffraction can be managed in a very broad frequency range. We introduce
optimized arrays of curved waveguides where light beams experience
{\em wavelength-independent} normal, anomalous, or zero diffraction.
Our results suggest novel opportunities for efficient
self-collimation, focusing, and reshaping of beams produced by {\em
white-light} and {\em super-continuum} sources. We also show how to
manipulate light patterns through {\em multicolor Talbot effect},
which is possible neither in free space nor in conventional photonic
lattices.
\end{abstract}

\pacs{42.25.Fx, 42.82.Et, 61.12.Bt}

\keywords{}

\maketitle

It is known that periodic photonic structures can be employed to
engineer and control the fundamental properties of light
propagation~\cite{Joannopoulos:1995:PhotonicCrystals,
Russell:1995-585:ConfinedElectrons}. In particular, the beam
refraction and diffraction can be modified dramatically, resulting
in many unusual phenomena. For example, a beam can experience {\em
negative refraction} in the direction opposite to normal at the
interface with a photonic
crystal~\cite{Russell:1995-585:ConfinedElectrons,
Notomi:2000-10696:PRB, Cubukcu:2003-604:NAT, Rosberg:2005-2293:OL}.
Additionally, the natural tendency of beams to broaden during
propagation can be controlled through diffraction
management~\cite{Eisenberg:2000-1863:PRL}. Diffraction can be
eliminated in periodic structures leading to self-collimation effect
where the average beam width does not change over hundreds of
free-space diffraction lengths~\cite{Rakich:2006-93:NAMT}. On the
other hand, diffraction can be made negative allowing for focusing
of diverging beams~\cite{Pertsch:2002-93901:PRL} and imaging of
objects with sub-wavelength resolution~\cite{Parimi:2003-404:NAT,
Lu:2005-153901:PRL}.

The physics of periodic photonic structures is governed by
scattering of waves from modulations of the refractive index and
their subsequent interference. This is a resonant process, which is
sensitive to both the frequency and propagation angle. Strong
dependence of the beam refraction on the optical wavelength known as
{\em superprism effect} was observed in photonic
crystals~\cite{Kosaka:1999-2032:JLT}. Spatial beam diffraction also
depends on the wavelength, and it was found in recent
experiments~\cite{Rakich:2006-93:NAMT, Longhi:2006-243901:PRL} that
the effect of beam self-collimation is restricted to a spectral
range of less than 10\% of the central frequency. Such a strong
dependence of the spatial beam dynamics on wavelength can be used
for multiplexing and demultiplexing of signals in optical
communication networks~\cite{Wu:2002-915:IQE, Wan:2005-353:OC}.
However, it remains an open question whether photonic structures can
be used to perform spatial steering and shaping of beams emitted by
white-light sources, such as light with supercontinuum frequency
spectrum generated in photonic-crystal fibers and fiber
tapers~\cite{Ranka:2000-25:OL, Wadsworth:2002-2148:JOSB}.

In this Letter, we suggest a novel type of periodic photonic
structures designed for {\em wavelength-independent diffraction
management} in a very broad frequency range, covering a spectral
range up to 50\% of the central frequency. We introduce the
optimized periodic structures where multicolor beams experience
constant normal, anomalous, or zero diffraction. This opens up novel
opportunities for efficient self-collimation, focusing, and shaping
of white-light beams. For example, in such optimized structures it
becomes possible to manipulate white-light patterns through {\em
multicolor Talbot effect}, which otherwise is not feasible in free
space or in conventional photonic lattices.

We study propagation of beams emitted by a continuous white-light
source in a periodic array of coupled optical waveguides [see
Fig.~\rpict{discreteDiffraction}(a)], where the waveguide axes are
also periodically curved in the propagation direction [see examples
in Figs.~\rpict{selfCollimation}(a)
and~\rpict{constantDiffraction}(a)]. In the linear regime, the
overall beam dynamics is defined by independent evolution of complex
beam envelopes $E(x,z;\lambda$) at individual frequency components
governed by the normalized paraxial equations,
\begin{equation} \leqt{nls}
   i \frac{\partial E}{\partial z}
   + \frac{z_s \lambda}{4 \pi n_0 x_s^2} \frac{\partial^2 E}{\partial x^2}
   + \frac{2 \pi}{\lambda} \nu\left[ x- x_0(z) \right] E = 0 ,
\end{equation}
where $x$ and $z$ are the transverse and propagation coordinates
normalized to the characteristic values $x_s= 1 \mu m$ and $z_s = 1
mm$, respectively, $\lambda$ is the vacuum wavelength, $c$ is the
speed of light, $n_0$ is the average refractive index of the medium,
$\nu(x) \equiv \nu(x+d)$ is the refractive index modulated with the
period $d$ in the transverse direction, and $x_0(z) \equiv x_0(z+L)$
defines the longitudinal bending profile of the waveguide axis with
the period $L \gg d$. When the tilt of beams and waveguides at the
input facet is less than the Bragg angle at each wavelength, the
beam propagation is primarily characterized by coupling between the
fundamental modes of the waveguides, and can be described by the
tight-binding equations taking into account the periodic waveguide
bending~\cite{Longhi:2005-2137:OL, Longhi:2006-243901:PRL},
\begin{equation} \leqt{dnls}
   i \frac{d \Psi_{n}}{d z}
   + C( \omega ) \left[\Psi_{n+1} + \Psi_{n-1}\right]
   = \omega \ddot{x}_0(z) n \Psi_{n},
\end{equation}
where $\Psi_n(z; \omega)$ are the mode amplitudes, $n$ is the
waveguide number, $\omega = 2 \pi n_0 d / \lambda$ is the
dimensionless frequency, and the dots stand for the derivatives.
Coefficient $C( \omega )$ defines a coupling strength between the
neighboring waveguides, and it characterizes diffraction in a
straight waveguide array with $x_0 \equiv 0$
\cite{Jones:1965-261:JOS, Somekh:1973-46:APL}. The coupling
coefficient decreases at higher
frequencies~\cite{Iwanow:2005-53902:PRL} and accordingly the beam
broadening is substantially weaker at shorter wavelengths, see
Figs.~\rpict{discreteDiffraction}(b-e).

\pict{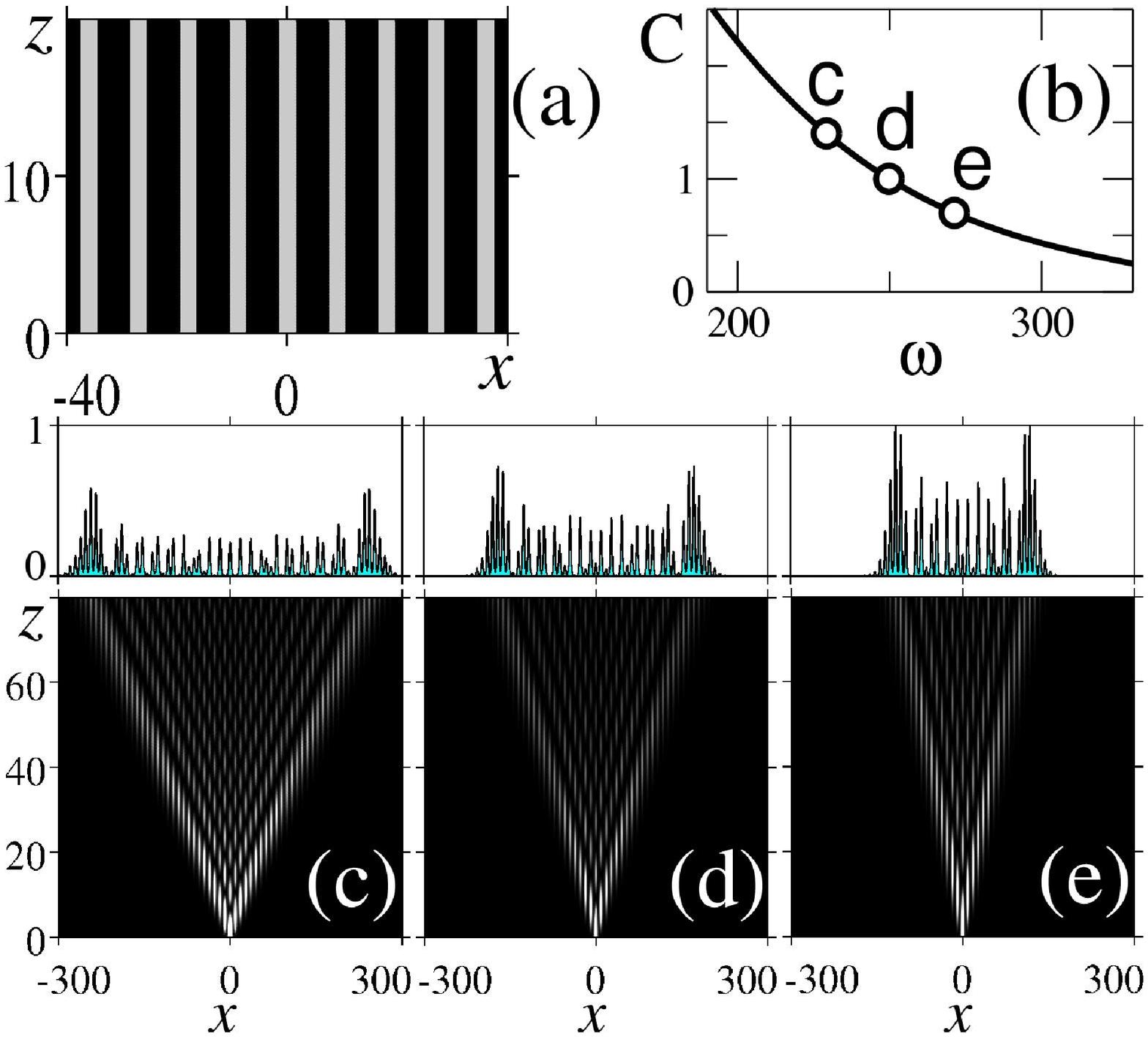}{discreteDiffraction}{Discrete diffraction
in (a)~straight waveguide array with period $d=9\mu m$. (b)~Coupling
coefficient normalized to the coupling at the central frequency
$C_0$. (c-e) Evolution of beam intensity and output intensity
profiles after $80mm$ propagation of a $3\mu m$ wide input beam for
(c)~$\lambda_r = 580nm$, (d)~$\lambda_0 = 532nm$, and (e)~$\lambda_b
= 490nm$, which correspond to the points `c', `d', and `e' in (b).
Waveguide width is $3\mu m$ and substrate refractive index is $n_0 =
2.35$. }

We consider symmetric profiles of the waveguide bending such that
${x}_0(z) = f(z - z_a)$ for a given coordinate shift $z_a$, where
function $f(z)$ is symmetric, $f(z) \equiv f(- z)$. Then, after a
full bending period ($z \rightarrow z+L$) the beam diffraction is
the same as in a straight waveguide array with the effective
coupling coefficient~\cite{Longhi:2005-2137:OL,
Longhi:2006-243901:PRL}
\begin{equation} \leqt{couplingC}
   C_{\rm eff}( \omega )
   = C( \omega ) L^{-1}
      \int_{0}^{L} \cos\left[ \omega \dot{x}_0(\zeta) \right] d\zeta .
\end{equation}
According to Eq.~\reqt{couplingC}, diffraction of multicolor beams
is defined by an interplay of bending-induced dispersion and
frequency dependence of the coupling coefficient in a straight
waveguide array. We suggest that spatial evolution of all frequency
components can be synchronized allowing for shaping and steering of
multi-color beams, when effective coupling remains constant around
the central frequency $\omega_0$,
\begin{equation} \leqt{constantC}
  \left. d C_{\rm eff}(\omega) / d \omega \right|_{\omega = \omega_0} = 0 ,
\end{equation}
and we demonstrate below that this condition can be satisfied by
introducing special bending profiles.

First, we demonstrate the possibility for {\em self-collimation of
white-light beams}, where all the wavelength components remain
localized despite a nontrivial evolution in the photonic structure.
Self-collimation regime is realized when the diffraction is
suppressed and the effective coupling coefficient vanishes, $C_{\rm
eff} = 0$. This effect was previously observed for monochromatic
beams in arrays with zigzag~\cite{Eisenberg:2000-1863:PRL} or
sinusoidal~\cite{Longhi:2006-243901:PRL} bending profiles, however
in such structures the condition of zero coupling cannot be
satisfied simultaneously with Eq.~\reqt{constantC}, resulting in
strong beam diffraction under frequency detuning by several
percent~\cite{Longhi:2006-243901:PRL}. We find that {\em broadband
diffraction management} becomes possible in hybrid structures with a
periodic bending profile that consists of alternating segments [see
example in Fig.~\rpict{selfCollimation}(a)], $x_0(z) =  A_1 \lbrace
\cos\left[2\pi z / z_0\right] - 1 \rbrace$ for $0 \leq z \leq z_0$,
$x_0(z) =  A_2 \lbrace \cos\left[2\pi (z - z_0) / (L/2 - z_0)\right]
- 1 \rbrace$ for $z_0 \leq z \leq L/2$, and $x_0(z) = - x_0(z-L/2)$
for $L/2 \leq z \leq L$. Effective coupling in the hybrid structure
can be calculated analytically, $C_{\rm eff}(\omega) = C(\omega) 2
L^{-1} [ z_0 J_0(\xi_1) + (L/2 - z_0)J_0(\xi_2) ]$, where $J_m$ is
the Bessel function of the first kind of the order $m$, $\xi_1 = 2
\pi A_1 \omega / z_0$, and $\xi_2 = 2 \pi A_2 \omega / \left(L/2 -
z_0\right)$.

We select a class of symmetric profiles of the waveguide bending to
avoid asymmetric beam distortion due to higher-order effects such as
third-order diffraction. Additionally, the waveguides are not tilted
at the input, i.e. $\dot{x}_0(z=0)=0$, in order to suppress
excitation of higher-order photonic bands by incident beams inclined
by less than the Bragg angle. The effect of Zener tunneling to
higher bands~\cite{Zener:1934-523:RAR, Trompeter:2006-23901:PRL} and
associated scattering losses can be suppressed irrespective of the
waveguide tilt inside the photonic structure by selecting
sufficiently slow modulation to minimize the curvature
$\ddot{x}_0(z)$ and thereby achieve adiabatic beam shaping.

\pict{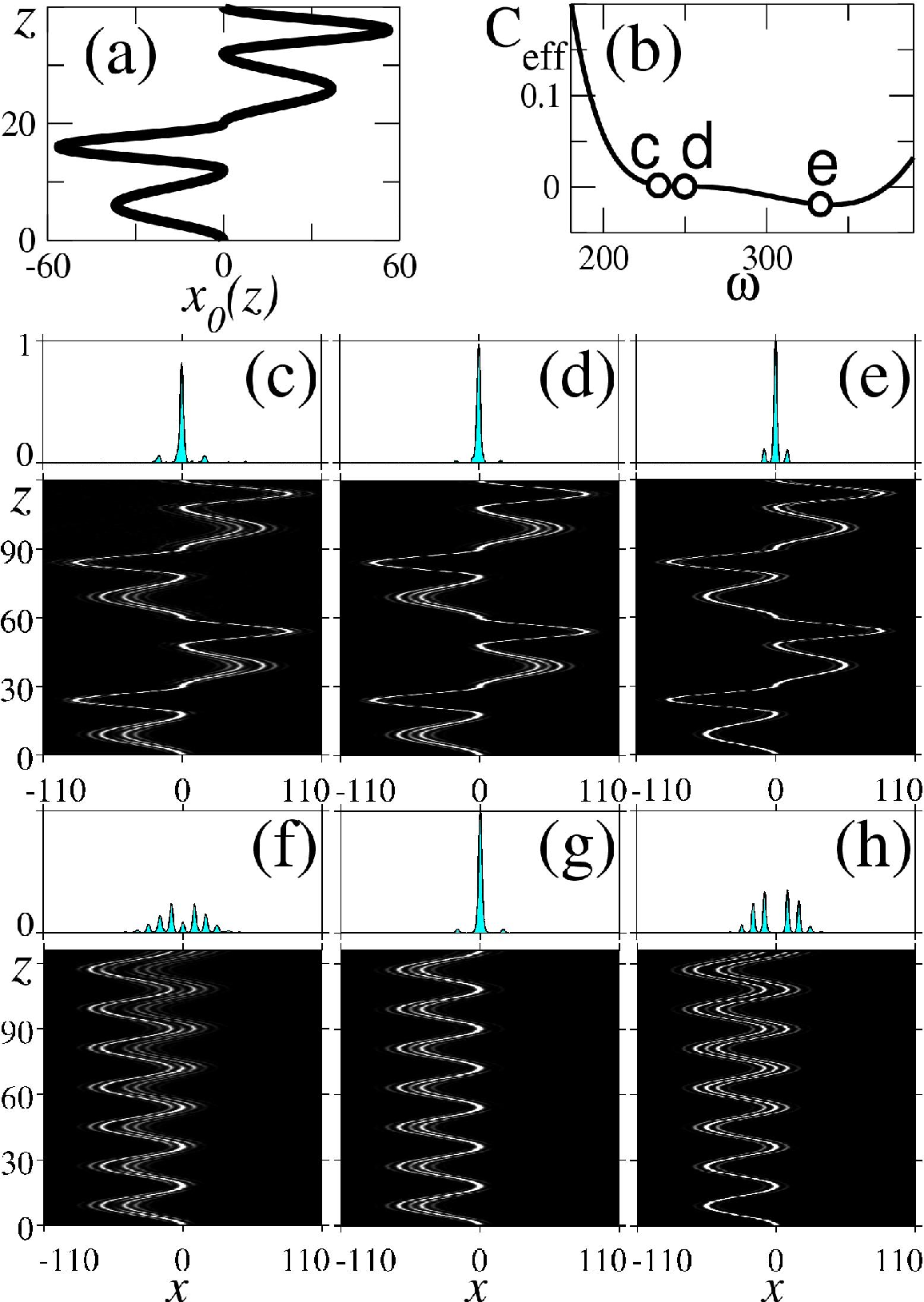}{selfCollimation}{ (a-e)~Broadband
self-collimation in an optimized waveguide array: (a)~Waveguide
bending profile with the period $L = 60mm$ and modulation parameters
$A_1 = 27\mu m$, $A_2 = 42\mu m$, $z_0 = 18mm$. (b)~Effective
coupling normalized to the coupling in the straight array at the
central frequency $C_0 = C(\omega_0)$. (c-e)~Evolution of the beam
intensity and output intensity profiles for different wavelengths
marked (c)~$\lambda_r = 560nm$, (d)~$\lambda_0=532nm$, and
(e)~$\lambda_b =400nm$ corresponding to marked points in (b).
(f-h)~Frequency-sensitive diffraction in array with the sinusoidal
bending profile at the wavelengths corresponding to plots~(c-e). }

In order to realize broadband self-collimation, we choose the
structure parameters such that $\xi_1(\omega_0) = \tilde{\xi_1}
\simeq 2.40$ and $\xi_2(\omega_0) = \tilde{\xi_2} \simeq 5.52$ are
the first and the second roots of equation $J_0(\tilde{\xi})=0$.
Then, the self-collimation condition is exactly fulfilled at the
central frequency $\omega_0$, $C_{\rm eff}(\omega_0) = 0$, and
simultaneously the condition of frequency-independent coupling in
Eq.~\reqt{constantC} is satisfied for the following modulation
parameters, $A_1 = [\tilde{\xi_1} \tilde{\xi_2} J_1( \tilde{\xi_2} )
       / 2 \pi (   \tilde{\xi_2} J_1(\tilde{\xi_2})
                 - \tilde{\xi_1} J_1(\tilde{\xi_1}) ) \omega_0] L / 2$,
$A_2 = - [ J_1(\tilde{\xi_1}) / J_1(\tilde{\xi_2}) ] A_1$, and $z_0
= 2 \pi \omega_0 A_1 / \tilde{\xi_1}$. As a result, we obtain an
extremely flat coupling curve shown in
Fig.~\rpict{selfCollimation}(b) where the point `d' corresponds to
the central frequency. In this hybrid structure not only the first
derivative vanishes according to Eq.~\reqt{constantC}, but the
second derivative vanishes as well, $\left| \left.d^2 C_{\rm
eff}(\omega) / d\omega^2\right|_{\omega = \omega_0} \right| \sim
\left| \tilde{\xi}_1 J_2(\tilde{\xi}_1) J_1(\tilde{\xi}_2) -
\tilde{\xi}_2 J_2(\tilde{\xi}_2)J_1(\tilde{\xi}_1) \right| <
10^{-15}$. As a result, the effective coupling remains close to
zero in a very broad spectral region of up to 50\% of the central
frequency. We note that the modulation period $L$ is a free
parameter, and it can always be chosen sufficiently large to avoid
scattering losses due to waveguide bending since the maximum
waveguide curvature is inversely proportional to the period, ${\rm
max} |\ddot{x}_0(z)| \sim L^{-1}$. Although the beam evolution
inside the array does depend on the wavelength, the incident beam
profile is exactly restored after a full modulation period, see
examples in Figs.~\rpict{selfCollimation}(c-e). Self-collimation is
preserved even at the red spectral edge, where coupling length is
the shortest and discrete diffraction in the straight array is the
strongest [cf. Fig.~\rpict{selfCollimation}(c) and
Fig.~\rpict{discreteDiffraction}(c)]. The hybrid structure provides
a dramatic improvement in the bandwidth for self-collimation effect
compared to the array with a simple sinusoidal modulation, where
beams exhibit diffraction under small frequency detuning, see
Figs.~\rpict{selfCollimation}(f-h).

\pict{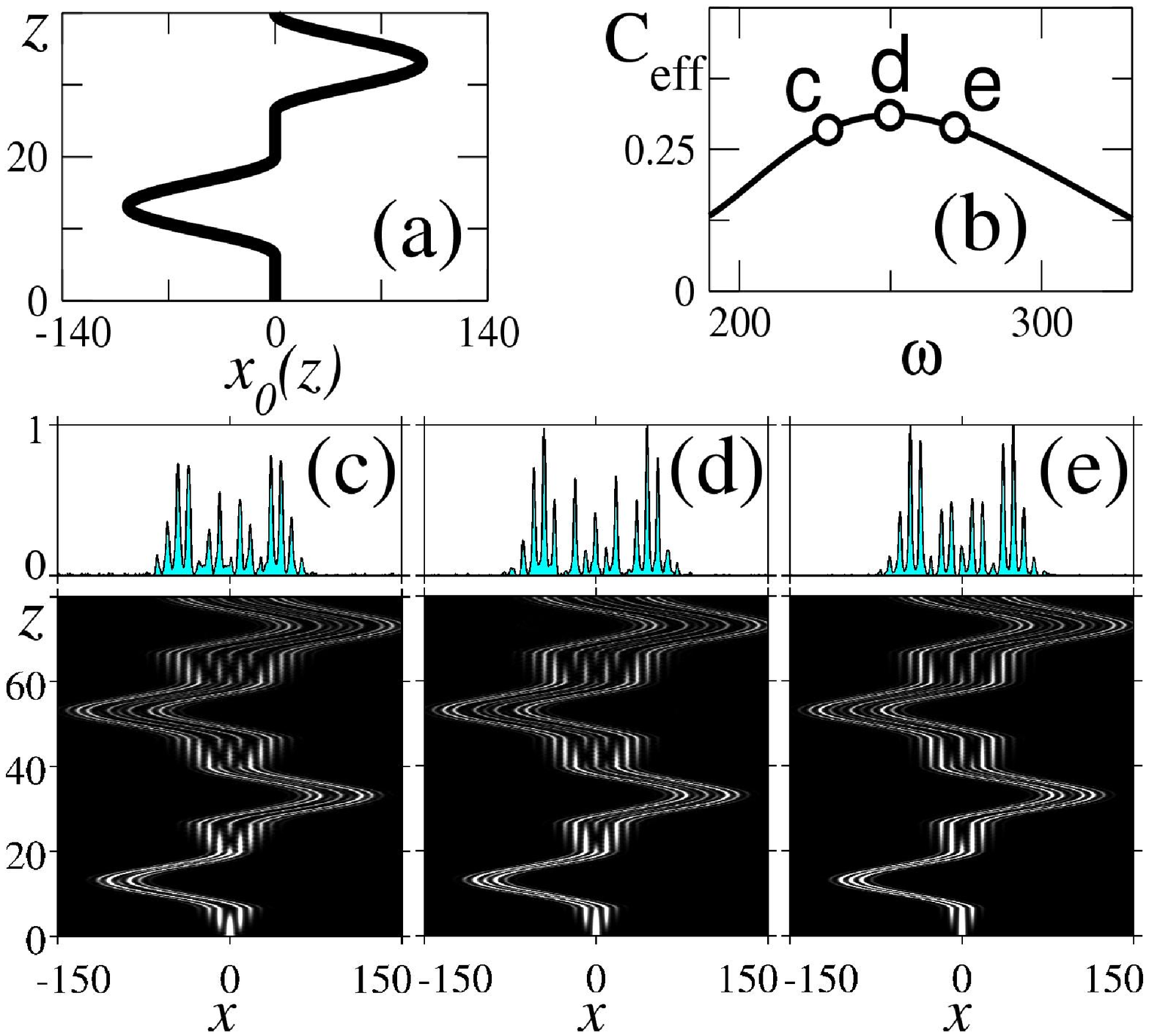}{constantDiffraction}{
Wavelength-independent diffraction in an optimized periodically
curved waveguide array. (a)~Waveguide bending profile with the
period $L = 40mm$ and (b)~corresponding effective coupling
normalized to the coupling in the straight array at the central
frequency $C_0 = C(\omega_0)$. (c-e)~Evolution of beam intensity and
output intensity profiles after propagation of two full periods for
the wavelengths (c)~$\lambda_r = 580nm$, (d)~$\lambda_0=532nm$, and
(e)~$\lambda_b =490nm$, which correspond to points `c', `d', and `e'
in plot~(b).}

We now analyze the conditions for {\em frequency-independent normal
or anomalous diffraction} that may find applications for reshaping
of multicolor beams. In order to reduce the device dimensions, it is
desirable to increase the absolute value of the effective coupling
and simultaneously satisfy Eq.~\reqt{constantC} to achieve broadband
diffraction management. We find that Eq.~\reqt{constantC} can be
satisfied even in the simplest two-segment hybrid structure with
$z_0 = L/2$ and $A_1 = \left(\xi / 2\pi\omega_0\right) L/2$. Here a
set of possible parameter values $\xi$ is determined form the
relation $J_0(\xi) / J_1(\xi) = C_0 \xi / C_1 \omega_0$, where $C_0
= C(\omega_0)$ and $C_1 = \left. d C(\omega) / d \omega\right|_{\omega
= \omega_0}$ characterize dispersion of coupling in a straight
array. It is possible to obtain both normal and anomalous
diffraction regimes for normally incident beams, corresponding to
positive and negative effective couplings $C_{\rm eff}(\omega_0) =
C_0 J_0(\xi)$ depending on the chosen value of $\xi$. For example,
for the waveguide array shown in Fig.~\rpict{discreteDiffraction},
at the central frequency $\omega_0 = 250$ [corresponding wavelength
is $\lambda_0 = 532nm$] coupling parameters are $C_0 \simeq
0.13mm^{-1}$ and $C_1 \simeq -0.0021mm^{-1}$. Then, constant
positive coupling around the central frequency $C_{\rm
eff}(\omega_0) \simeq  0.25 C_0$ is realized for $\xi \simeq 6.47$
and constant negative coupling $C_{\rm eff}(\omega_0) \simeq  -0.25
C_0$ for $\xi \simeq 2.97$.

We perform a comprehensive analytical and numerical analysis, and
find that a hybrid structure with bending profile consisting of one
straight (i.e $A_1 \equiv 0$) and one sinusoidal segment can provide
considerably improved performance if $\omega_0 C_1 / C_0 >
\xi_{cr}J_1(\xi_{cr}) / J_0(\xi_{cr})$, where value $\xi_{cr} \simeq
5.84$ is found from the equation $\left[J_1(\xi_{cr}) +
\xi_{cr}\left[J_0(\xi_{cr}) - J_2(\xi_{cr})\right] /
2\right]\left[J_0(\xi_{cr}) - 1\right] + \xi_{cr}J_1^2(\xi_{cr}) =
0$. Under such conditions, larger values of positive effective
coupling can be obtained in a hybrid structure with $A_1 \equiv 0$,
$A_2 = [C_1 C_{\rm eff}(\omega_0) / 2\pi C_0^2J_1(\tilde{\xi_2})]
L/2$, $z_0 = [C_{\rm eff}(\omega_0) / C_0] L/2$. In this structure,
the effective coupling at central frequency is $C_{\rm
eff}(\omega_0) = \tilde{\xi_2}C_0^2J_1(\tilde{\xi_2}) /
[\tilde{\xi_2}C_0J_1(\tilde{\xi_2}) + \omega_0C_1]$.

Example of a hybrid structure which provides strong
wavelength-independent diffraction is shown in
Fig.~\rpict{constantDiffraction}(a), and the corresponding effective
coupling is plotted in Fig.~\rpict{constantDiffraction}(b). The
output diffraction profiles in this optimized structure are very
similar in a broad spectral region, see examples for three
wavelengths in Figs.~\rpict{constantDiffraction}(c-e). We note that
the outputs at these wavelengths are substantially different after
the same propagation length in the straight waveguide array, as
shown in Figs.~\rpict{discreteDiffraction}(c-e).

As one of the applications of the broadband diffraction management
we consider a {\em multicolor Talbot effect} which allows to
manipulate white-light patterns. The Talbot effect, when any
periodical monochromatic light pattern reappears upon propagation at certain
equally spaced distances, has been known since the famous discovery
in 1836~\cite{Talbot:1836-401:RAR}. It was recently shown that the Talbot effect is also
possible in discrete systems for certain periodic input
patterns~\cite{Iwanow:2005-53902:PRL}. For example, for the
monochromatic periodic input pattern of the form
$\{1,0,0,1,0,0,\ldots\}$, Talbot revivals take place at the distance
$L_T^{(1)} = \left(2\pi / 3\right)\left[1 / C(\omega)\right]$, see
Fig.~\rpict{Talbot}(a).

\pict{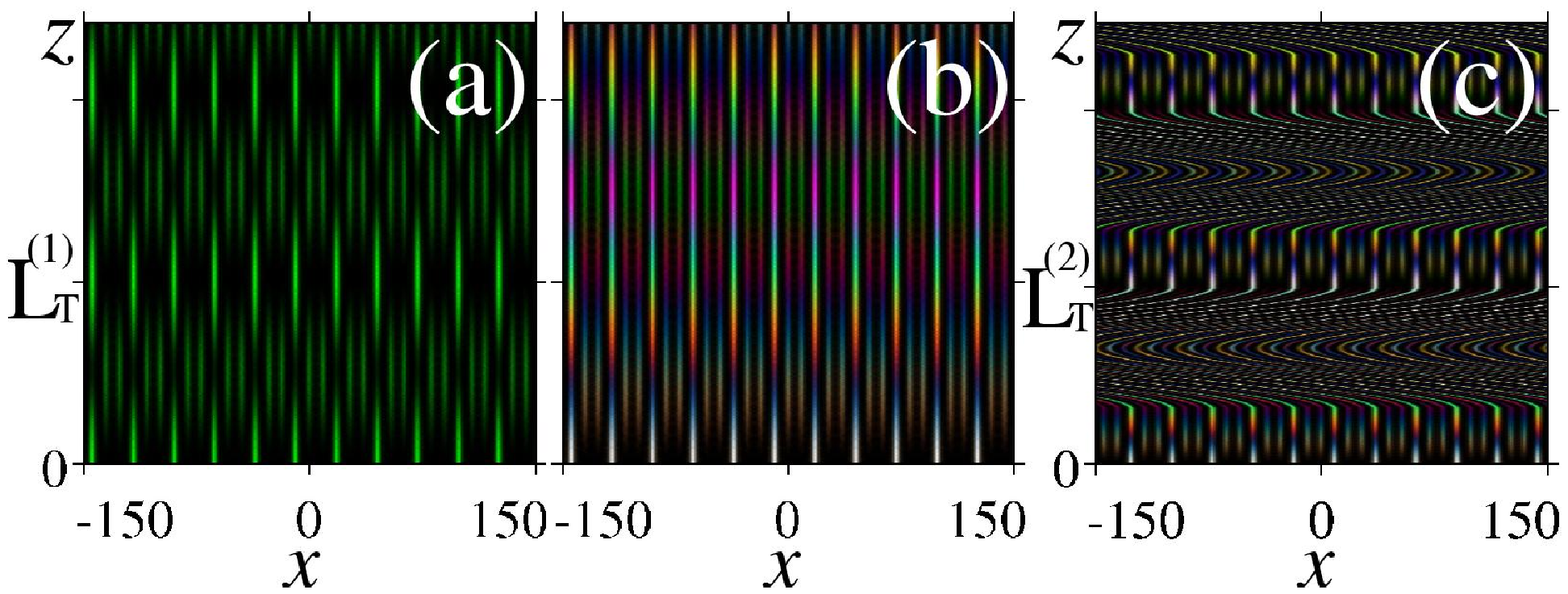}{Talbot}{ (a)~Monochromatic Talbot effect in the
straight waveguide array shown in
Fig.~\rpict{discreteDiffraction}(a): periodic intensity revivals
every $L_T^{(1)} = 16.5mm$ of propagation for the input pattern
$\{1,0,0,1,0,0,\ldots\}$ and the wavelength $\lambda_0=532nm$.
(b)~Disappearance of the Talbot carpet in the straight array when
input consists of three components with different wavelengths
$\lambda_r = 580nm$, $\lambda_0=532nm$, and $\lambda_b =490nm$ and
equal intensities. (c)~Multicolor Talbot effect in the optimized
structure with wavelength-independent diffraction [see
Fig.~\rpict{constantDiffraction}.] Half of the bending period $L/2 =
L_T^{(2)} = 53.2mm$ is equal to the Talbot distance for the
corresponding effective coupling length. }

Period of the discrete Talbot effect in the waveguide array is
inversely proportional to the coupling coefficient $C(\omega)$,
which strongly depends on frequency, see
Fig.~\rpict{discreteDiffraction}(b). Therefore, for each specific
frequency Talbot recurrences occur at different
distances~\cite{Iwanow:2005-53902:PRL}, and periodic intensity
revivals disappear for the multicolor input, see
Fig.~\rpict{Talbot}(b). Multicolor Talbot effect is also not
possible in free space where revival period is proportional to
frequency. Most remarkably, multicolor Talbot effect can be observed
in optimized waveguide arrays with wavelength-independent
diffraction, see Fig.~\rpict{Talbot}(c). In this example, we use the
shape of structure with constant positive diffraction shown in
Fig.~\rpict{constantDiffraction}, and choose half of the bending
period to be equal to the period of the Talbot recurrences for the
corresponding effective coupling in this structure, $L_T^{(2)} =
\left(2\pi / 3\right)\left[1 / C_{\rm eff}(\omega)\right]$. We note
that the length of the straight segment is equal to the Talbot
distance at central wavelength in the straight array, $z_0 =
L_T^{(1)}$, which explains partial revivals at the end of the
straight segments visible in Fig.~\rpict{Talbot}(c).

In conclusion, we have introduced a novel class of photonic
structures where diffraction can be engineered in a very broad
frequency range. We have analyzed the optimized array of
periodically curved waveguides where light beams experience
wavelength-independent normal, anomalous, or zero diffraction, and
predicted the multicolor Talbot effect which is not possible in free
space and conventional waveguide arrays. Our results suggest novel
opportunities for efficient self-collimation, focusing, and
reshaping of beams produced by white-light and super-continuum
sources.

\end{sloppy}
\end{document}